\documentclass{iopart}
\usepackage{iopams}
\usepackage{graphicx}
\usepackage{bm}
\begin{document}

\title[Effect of chemical disorder on NiMnSb investigated by Appearance Potential Spectroscopy]{Effect of chemical disorder
on NiMnSb investigated by Appearance Potential Spectroscopy: a theoretical study}

\author{J Min\'ar$^{1}$, J Braun$^{1}$, S Bornemann$^{1}$, H Ebert$^{1}$ and M Donath$^{2}$}

\address{$^1$ Department Chemie und Biochemie, Lehrbereich Physikalische Chemie, Ludwig-Maximilians-Universit\"at 
M{\"u}nchen, Butenandtstr 5-13, 81377 M\"unchen, Germany}
\address{$^2$ Physikalisches Institut, Westf\"alische
Wilhelms-Universit{\"a}t M{\"u}nster, Wilhelm-Klemm-Str.~10, 48149
M{\"u}nster, Germany}

\ead{jan.minar@cup.uni-muenchen.de}

\date{\today}

\begin{abstract}
The half-Heusler alloy NiMnSb is one of the local-moment ferromagnets with unique properties for future
applications. Band structure calculations predict exclusively majority bands at the Fermi level,
thus indicating {100\%} spin polarization there. As one thinks about applications and the design of functional
materials, the influence of chemical disorder in these materials must be considered. The magnetization, spin
polarization, and electronic structure are expected to be sensitive to structural and stoichiometric changes.
In this contribution, we report on an investigation of the spin-dependent electronic structure of NiMnSb.
We studied the influence of chemical disorder on the unoccupied electronic density of states by use of the
ab-initio Coherent Potential Approximation method. The theoretical analysis is discussed along with
corresponding spin-resolved Appearance Potential Spectroscopy measurements. Our theoretical approach
describes the spectra as the fully-relativistic self-convolution of the matrix-element weighted, orbitally
resolved density of states. 
\end{abstract}

\submitto{\JPD}

\section{Introduction}\label{intro}
In 1903, first Heusler alloys (HA) have been singled out as a special group of magnetic materials that
exhibit ferromagnetism in compounds of non-magnetic elements \cite{Heusler1}. Currently, the Mn-based
half-Heusler alloys with the generic formula $X$Mn$Z$ (with $X$ being a $3d$ metal and $Z$ belonging
to the III or IV group) are well known magnetic systems that crystallizes in the Heusler $C1_b$ structure,
which is closely related to the zincblende structure. Especially, the half-metallic ferromagnet 
NiMnSb \cite{degroot83} has been triggered a variety of experimental and theoretical investigations
due to the search for materials to be used as spin-injectors. These materials are defined as magnetic
materials with a band gap at the Fermi level for electrons of one spin direction. Band structure
calculations for bulk NiMnSb show a gap of about 0.5 eV at T=0 for the minority electrons, which means
{100\%} spin polarization at the Fermi level $E_{\rm F}$. In fully relativistic calculations a reduced
but not vanishing gap and {99\%} spin polarization at $E_{\rm F}$ are predicted \cite{mavropoulos04}.
As in the case of the group III-V semiconductors, the crystal
structure is the reason for the band gap. 
It is, therefore, called a covalent band gap \cite{fang02}. As a consequence, the crystal structure
and the site occupation within the given structure are important for the appearance of the gap. Atomic
disorder, especially interchange of atoms between the Ni and Mn sub-lattices, results in a strong reduction of the spin polarization at
$E_{\rm F}$ \cite{orgassa99}. However, this interchange costs approximately as much energy as the
evaporation of the metallic constituents \cite{fang02}. Therefore, half-metallic behavior is an
idealization distorted by structural inhomogeneities \cite{SpinPolarizationOfmaterials,55percentatCo2MnSi,Brown2000}
and through the decrease of the spin polarization of spin carriers and vanishing of the semiconducting
band gap for minority spin electrons in real crystals and interfaces caused by finite temperature
effects.

A variety of experimental results for NiMnSb is available in the literature. While bulk measurements
support the half-metallic behavior (see, e.g., \cite{vanderheide85,hanssen90}), surface-sensitive
techniques have failed so far to detect the energy gap for minority electrons or the {100\%} spin
polarization at the Fermi level $E_{\rm F}$. Most of the studies were performed on polycrystals or thin
films, single and polycrystalline, prepared under various growth conditions on a variety of substrates.
Spin-integrated photoemission (PE) spectra obtained with a photon energy of 45 eV showed clear signatures
of the Ni and Mn $3d$ contributions \cite{robey92}. Angle-integrated PE with variable photon energies
identified Ni and Mn $3d$ emission from the valence bands emphasized the need for further careful
investigations both experimentally and theoretically \cite{kang93}. A spin-polarized PE study found
values of the spin polarization up to {50\%} near photothreshold \cite{bona85}. The authors concluded that,
if an energy gap for minority electrons should exist, it is smaller than 0.5 eV. A further spin-resolved PE
work, yet with photon energies between 38 and 76 eV, reported spin polarization values of at most {40\%}
close to the Fermi level \cite{zhu01}. A reduced surface magnetization at remanence and/or a surface phase
different from the bulk were proposed as possible explanations for the unexpected low polarization values.
Furthermore, an angle-resolved photoemission study of the electronic structure single-crystalline NiMnSb
sample showed that the binding energies of the photoemission intensities are sensitive to the particular
surface condition and/or preparation \cite{Cor2006}. Additionally, a
possible surface state has been found
in these $\bf k$-resolved measurements. In another study, a temperature-dependent cross-over from
half-metallic to normal ferromagnetic behavior at 80 K was reported and discussed as a possible reason for
the detected low polarization \cite{hordequin00}. It was suggested that, at room temperature, the spin
polarization at $E_{\rm F}$ may be considerably lowered due to a populated minority band, although the
magnetization is not much reduced compared to the value at T=0. (With a Curie temperature of about 730 K,
the magnetization at room temperature amounts to $\approx$ {92\%} of the saturation value \cite{ritchie03}.)
However, even point contact Andreev reflection measurements at 4.2 K at the free surface of NiMnSb gave a
maximal value of only {44\%}, independent of different surface preparations and magnetic domain structures
\cite{cloves04}. This result is in line with former spin-polarized PE data, which did not find a higher
polarization value for 20 K than for room temperature \cite{bona85}. As a complement to the reports above,
a room-temperature study by angle and spin-resolved inverse photoemission reported close to {100\%} spin
polarization at $E_{\rm F}$ and $\overline\Gamma$ under some conditions \cite{ristoiu00}. Last but not
least from a spin-resolved Appearance Potential Spectroscopy (APS) study the measured spin asymmetry from
the surface region was found to be significantly reduced compared to the theoretical prediction \cite{Kolev2005}.

In this contribution, we present spin-integrated and spin-resolved APS spectra from NiMnSb(001) and discuss
by comparison with experimental data the degree of spin polarization as a function of chemical disorder.
The paper is organized as follows. Section 2 is devoted to some experimental and computational details.
In section 3 we discuss the theoretical results and compare with corresponding experimental data. A summary
is given in section 4.

\section{Experimental and computational details}
\subsection{Spectroscopical scheme}
In this section the relevant details of the experimental
investigations are given. A more delayed report can be found in Ref.\ \cite{Kolev2005}.

APS is a surface sensitive tool to study the unoccupied density of states (DOS) with elemental resolution
\cite{park72}. For spin-resolved APS, the sample is bombarded with a spin-polarized electron beam of variable
energy while the total yield of emitted x-rays or electrons is monitored \cite{ertl93,reinmuth97}. At energies
high enough to excite a core electron into empty states above the Fermi level, the yield of emitted particles
increases due to recombination of the created core hole via x-ray or Auger electron emission. In our case, we
detect the emitted x-rays \cite{rangelov98}. Potential modulation together with lock-in techniques are used
to separate the small signal from the dominating background. Since both the exciting and the excited electron
are scattered into empty states, the rate of possible excitations and, thereby, of detected recombinations
depends on the DOS above $E_{\rm F}$. The spin-polarized electron beam used for excitation is emitted from a
GaAs photocathode irradiated with circularly polarized 830 nm laser light. This arrangement provides about
30\% spin polarization of the emitted electrons \cite{kolac88}. The spectra shown have been renormalized
to 100\% hypothetical beam polarization. The APS signal depends on the spin of the exciting electron in the
case of ferromagnets because of the spin-dependent unoccupied DOS. The elemental resolution comes from the
fact that core levels are involved whose energies are characteristic of the various elements.

\subsection{Calculational scheme}
 
\begin{figure}
\centering
\includegraphics[height=10cm]{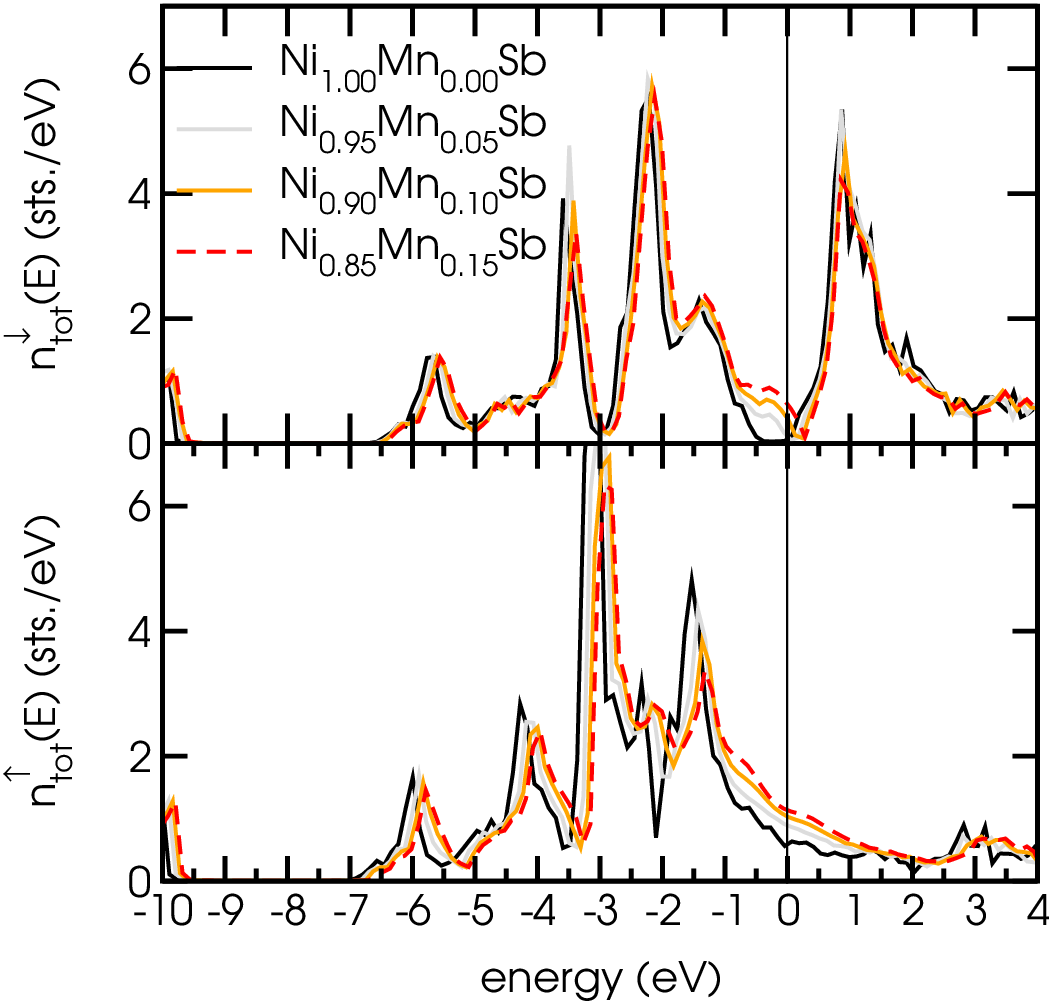}
\caption {Spin-resolved total DOS for NiMnSb as a function of chemical disorder. The gap in the minority
spin channel closes with increasing disorder induced by interchange of Mn and Ni atoms}
\end{figure}

Hence, the calculational procedure concerning APS is found in detail
in \cite{Kolev2005,Don2005} we restrict 
ourselves on a short description of some computational details. The
fully relativistic DOS calculations have 
been performed within the SPRKKR computer program \cite{ebekkr}
running in the CPA-mode \cite{ebe00}. 
The chemical disorder introduced in the spectroscopical analysis
consists of an interchange between Mn 
atoms on Ni sites and vice versa with concentration values x of 0$\%$,
5$\%$, 10$\%$ and 15$\%$. This 
means we have calculated (Ni$_{1-x}$Mn$_{x}$)(Mn$_{1-x}$Ni$_{x}$)Sb in
the L2$_1$ crystal structure using 
the experimental lattice constant 5.91 $\AA$. In Figs.~2 and 3 we use
the compact notation Ni$_{1-x}$Mn$_{x}$Sb. 
Figure~1 shows the spin-resolved total DOS for NiMnSb. Clearly visible
is the closing of the spin-dependent 
gap at the Fermi energy caused by the chemical disorder. Even at 5$\%$ interchange of Mn and Ni atoms
the gap completely vanishes because the spectral weight is shifted towards
the Fermi level. A significant enhancement of the spectral weight around the Fermi level is also observable in
the majority DOS and therefore affect the amount of spin polarization that is measured in an APS experiment.
The corresponding partial densities of states together with the spin-dependent potentials serve as input
quantities for the spectroscopical calculations. In our case we probed the local DOS (LDOS) at the Mn atom
in NiMnSb. The surface sensitivity is a result of the short mean free path of the electrons in the solid for
energies that correspond to the Mn$_{2p}$ excitation ($\approx$ 640-650 eV). The elemental resolution results
from the fact that a core level with an element-specific binding energy is involved. Due to the excitation with
spin-polarized electrons the APS method becomes a magnetically sensitive technique. The calculated spin-dependent
APS-intensities result from the weighted self-convolution of the corresponding LDOS. The weighting factor itself
is given by the spin-dependent effective cross section consisting of
an appropriate combination of relativistic Coulomb matrix
elements. For a quantitative comparison between APS measurements and theoretical spectra one has to take into
account various lifetime effects and the experimental broadening. Electron lifetime effects have been included
in our analysis in a phenomenological way introducing a parametrized complex inner potential with an imaginary
part of V$_{0i}$ = 30 eV. The core hole lifetime is set to zero. Lifetime effects in the valence-band states are
accounted via a Lorentzian with an energy-dependent width. For the explicit parametrization of the Lorentzian
the reader is referred again to \cite{Kolev2005}. Strong correlations manifest themselves in form of the simple
scaling factor $\lambda$ = 0.7(E-E$_F$). Furthermore, due to experimental conditions the first derivative of the
APS signal has to be calculated and the apparatus broadening is considered by convoluting the calculated spectra
by a Gaussian of FWHM = 1.4 eV.
 
\begin{figure}
\centering
\includegraphics[height=12cm]{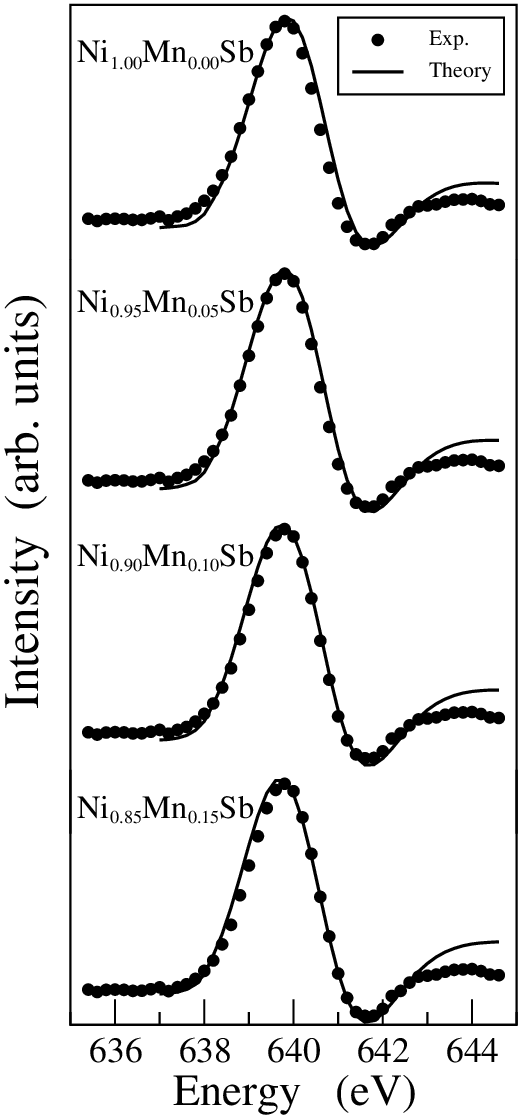}
\caption {Spin-integrated AP spectra for NiMnSb(001) at the Mn 2p$_{3/2}$
threshold as a function of chemical disorder. Experimental data are given by filled circles}
\label{fig2don}
\end{figure}

\section{Results and Discussion}
Figure~2 presents calculated spin-integrated AP spectra for NiMnSb(001) in comparison with APS data. The spectra
are normalized to equal maximum intensity. The theoretical spectra have been calculated with respect to the Fermi
level. To be comparable with the experiment the calculated spectra were shifted along the energy axis until
the peak maxima of the measured and calculated spectra coincidence. As it can be seen good agreement between
experiment and theory is achieved. The almost quantitative agreement suggests that the DOS as calculated resembles
the Mn local DOS in NiMnSb quite well in the spin-integrated case. The spectra show a pronounced structure at
639.7 eV from the Mn 2$p_{3/2}$ threshold. These feature originate from the self-convoluted local density of
unoccupied $3d$ states. The spectral feature at the high-energy side of the main line corresponds to the maximum of
the $sp$-like DOS, whose appearance and energetic position is sensitive to the short-range crystallographic order
around the atom where the local excitation occurs. As expected, the effect of chemical disorder is nearly negligible
concerning the spin-integrated DOS. Only slight deviations in the APS line shapes between the spectra calculated
for the ordered and for the disordered structures are observable. 

\begin{figure}
\centering
\includegraphics[height=12cm]{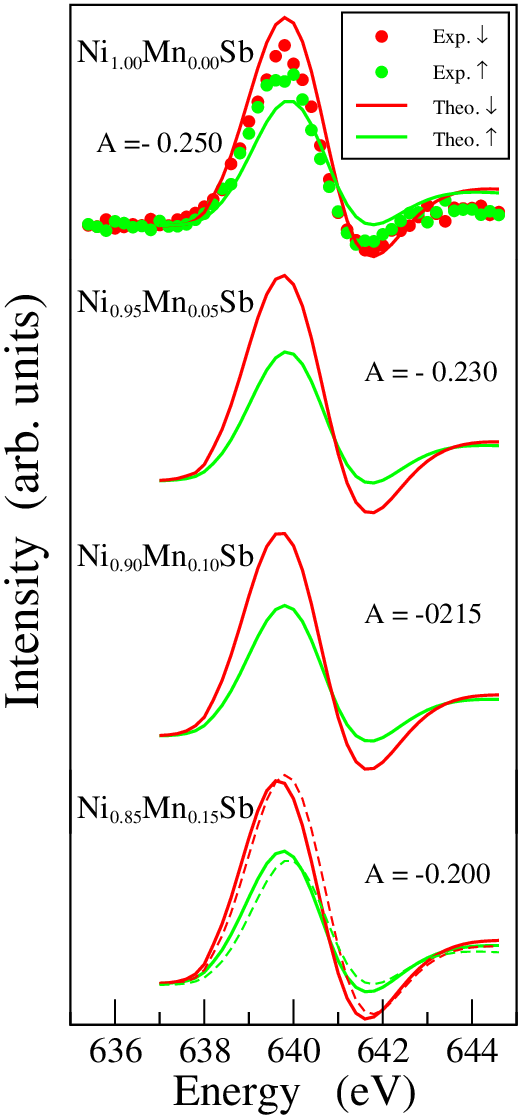}
\caption {Spin-resolved AP spectra for NiMnSb(001) at the Mn 2p$_{3/2}$
threshold as a function of chemical disorder. Experimental data are given by filled circles}
\label{fig3don}
\end{figure}

By adding spin resolution (see spectra in Fig.~3), we are able to obtain information about the spin-dependent DOS.
The experimental data for majority ($\uparrow$) and minority ($\downarrow$) spins are presented by black and grey
open circles, respectively. First of all, the APS lines show a clear spin asymmetry, $A$, between the spin-dependent
intensities $I_\uparrow$ and $I_\downarrow$ [$A=(I_\uparrow-I_\downarrow)/(I_\uparrow+I_\downarrow)$].
The experimental spin-resolved spectra of Fig.~3 represent an average from many different surface regions.
By taking into account the reduced remanent magnetization \cite{Kolev2005}, we end up with an estimated spin
asymmetry of $A=-0.115\pm0.012$ for a NiMnSb sample with saturated magnetization. The negative asymmetry reflects
the high density of unoccupied minority states in NiMnSb, i.e. the unoccupied minority Mn $3d$ states. The
theoretical spectra calculated for the ordered structure, however, predict a higher negative spin asymmetry in
the main APS peak of $A=-0.250$. This discrepancy between experiment and theory by a factor of two and more is
in line with the findings of earlier spin-resolved electron spectroscopic experiments. Introducing chemical
disorder the calculated value for the spin asymmetry decreases. This has to expected because of the pronounced
increase of spectral weight in the majority spin DOS around the Fermi energy that overcompensates the shift of
spectral weight in the minority spin DOS. The concentration dependence of the spin asymmetry is demonstrated by
the series of spin-dependent APS data shown in Fig.~3. The spin asymmetry in the main peak is calculated to
$A=-0.230$ for 5$\%$ interchange of Ni and Mn atoms and decreases from $A=-0.215$ for 10$\%$ disorder to
$A=-0.200$ for 15$\%$ interchange of Ni and Mn atoms. As a guide for the eye we compare at the bottom
of Fig.~3 directly the spectra which belong to zero and 15$\%$ disorder to clarify the decrease in the spin
asymmetry. Therefore, our analysis supports the result found by
Orgassa et al. that chemical disorder significantly lowers the spin polarization at $E_{\rm F}$. On the other
hand, introducing the effect of chemical disorder to our APS calculations, we end up with an improved agreement
between experiment and theory but the predicted spin asymmetry is still too high by nearly a factor of two. Our
analysis indicates that other mechanisms besides chemical disorder must be taken into account for a quantitative
explanation of the strongly reduced spin polarization in NiMnSb(001). 
 
\section{Summary}
We have presented spin-integrated and spin-resolved APS data for the unoccupied electronic states of NiMnSb as
a function of chemical disorder. The calculated series of spin-resolved and concentration dependent APS spectra 
demonstrates that chemical disorder influences the spin asymmetry of the unoccupied states
of NiMnSb. Concerning the spin polarization at the Fermi level, one must note that APS is not particularly sensitive
in that region. The nature of the APS signal doesn't allow one to resolve small energy gaps of the order of 0.5 eV
giving rise to a positive spin asymmetry at $E_{\rm F}$. The APS signal is rather dominated by the minority $d$
states above $E_{\rm F}$ leading to a negative spin asymmetry. On the other hand the significantly reduced 
spin polarization observed in the experiment clearly indicates a uncomplete spin polarization with a value
well below 100\%~. Therefore, the expected 100\%~spin polarization at the Fermi level is not verified, as in a
number of spin-polarized electron spectroscopic experiments before. Furthermore, one has to take into account
other mechanisms than chemical disorder to be able to explain quantitatively the unexpected low value for the
measured spin asymmetry. Therefore, we conclude that the following issues must be examined in future studies
on well-defined samples. First the magnetization within the surface layers compared with the bulk should be
investigated in more detail. Also surface phases with respect to composition and/or crystallographic order
different from the bulk could be important and the position and width of the band gap should be analyzed in more
detail in the calculations. A first study on NiMnSb(110) showed the formation of microstructures upon preparation,
which has consequences for the magnetic properties of the surface \cite{eick2007}. Even with a stoichiometric
surface, the number of nearest neighbors is changed at the surface and surface/interface states may form and
influence the gap \cite{jenkins01,jenkins02}. In the case that the free surface will not provide 100{\%} spin
polarization, there is still hope that specific interfaces may open the way for 100{\%} spin-polarized
charge-injection \cite{wijs01} in spintronic devices. Other materials with half-metallic behavior must be
considered as well in order to design functional materials for future applications \cite{fonin04}.

\end{document}